# Geometrical considerations in the control and manipulation of conductive heat flux

**Krishna P. Vemuri** and **P. R. Bandaru**,

Department of Mechanical and Aerospace Engineering,

University of California, San Diego, La Jolla, CA 92093, USA

**Abstract**:

We indicate the fundamental rationale underlying the control of temperature and the manipulation of thermal flux, with reference to a multilayered composite material. We show that when the orientation of the layers in the composite is physically rotated with respect to a constant temperature gradient, there would then be a corresponding introduction of off-diagonal components in the thermal conductivity tensor and thermal anisotropy is induced. The consequent bending of the heat flux lines is found to depend on both the (i) composite rotation angle, as well as the (ii) ratio of the thermal conductivities of the constituent materials.

**Text:**

The control of heat flux, through the rational design and arrangement of materials could form the basis for the creation of novel elements aimed at channeling thermal energy, *e.g.,* through either concentration or cloaking of the heat flux [1]. Devices based on such elements could find widespread use in various applications incorporating portable electronics and microprocessors, heat recovery from exhaust gas, integrated micro-combustion systems, battery devices, heat sinking modules in electronic devices, enhanced efficiencies for solar thermal energy utilization, *etc*. Consequently, there would be substantial progress towards the long cherished objectives of reducing energy loss and controlling heat propagation.

While fundamental concepts such as (a) the thermal extremum principle - where the propagation of heat takes the path of least thermal resistance [2], as well as the use of (b) coordinate transformation techniques [3,4]– for inducing anisotropy in the thermal conductivity ($\kappa$) have been proposed earlier, for the control of heat, a practical basis for the assembly of such elements as well as their limitations has not been well explored. Recently, a few experimental implementations [5] have indicated such possibilities; however, a clear analytical understanding has not been adequately achieved, and the clarification of the underlying issues constitutes the major aim of this paper. Much initial work in heat flux control had also been initially motivated from principles formulated for electromagnetic waves and subsequently adapted to heat transport, *e.g.,* related to transformation optics [6–8]. While relevant and interesting analogies do exist, *e.g.,* in two dimensions, there could be an equivalence between acoustics (applicable to very long wavelength phonons/heat transport) and electromagnetics in isotropic media [9], the relationship is not very clear. Moreover, the symmetry of the Maxwell equations for electromagnetic waves is not apparent in the Fourier law of heat conduction [10,11], where heat transport is diffusive [12,13] with the flux in the $i^{th}$ direction ($q_i$):

$$q_i = -\kappa_{ij} \nabla T_j \tag{1}$$

The $\kappa_{ij}$ represent the components of the second order thermal conductivity tensor, with respect to a rectangular (*x-y-z*) coordinate system, *i.e.,*

$$\kappa_{ij} = \begin{pmatrix} \kappa_{xx} & \kappa_{xy} & \kappa_{xz} \\ \kappa_{yx} & \kappa_{yy} & \kappa_{yz} \\ \kappa_{zx} & \kappa_{zy} & \kappa_{zz} \end{pmatrix} \tag{2}$$

with $\nabla T_j$ being the temperature gradient in the $j^{th}$ direction (*i=j,* for materials with isotropic thermal conductivity). However, engineering practice, to date, has been mostly focused on materials with isotropic thermal conductivity [14], [15], *i.e.,* $\kappa_{ij}$ is equated to a single scalar value ($\equiv \kappa$, say), and reported as such. In this case, the heat flux density vector (***q**_i*) follows the respective temperature gradient, *i.e.,* the heat flux in the horizontal/*x*-direction is only determined by the temperature gradient in that direction. The presence of off-diagonal terms, *i.e.,* $\kappa_{ij}$ with $i \neq j$, would induce cross-coupling and concomitant *bending* of the heat flux, *e.g.,* with a substantial $\kappa_{xy}$, the heat flux in the *x*-direction would be determined by the temperature gradient in *both* the *x*- and an orthogonal/*y-/z-* direction. Such considerations aimed toward the controlled bending/manipulation of the thermal flux lead to the study of materials with anisotropic thermal conductivity, where the off-diagonal components of $\kappa_{ij}$ would be crucial.

It was noted that the anisotropy needed for the bending of the heat flux could be obtained either by (a) having a material with anisotropic values of the thermal conductivity, or (b) by simply layering the materials, *e.g.,* when two sheets with nominally *isotropic* thermal conductivities $\kappa_1$ and $\kappa_2$ are alternatively stacked – as depicted in Fig. 1. (The $\kappa_1/\kappa_2$ ratio is assumed to be greater than unity, as layers 1 and 2 are inter-changeable in our formulation. Consequently, the material with the higher thermal conductivity should be taken to have a value of $\kappa_1$. The case of $\kappa_1/\kappa_2 = 1$ corresponds to a homogeneous material). As materials conforming to (a) are relatively rare [15], we focus on (b), and show that such a layered configuration can be modified through geometrical considerations, and would enable a tuning of the effective thermal conductivity and conductance. Considering one-dimensional parallel and perpendicular heat transport, between the surface on the right (maintained at a temperature: $T_h$) and the left surface (at a lower temperature: $T_c$), assuming flux continuity and neglecting interfacial effects, we can easily derive (as in the Supplementary Information, section S1):

$$\kappa_{ij} = \begin{pmatrix} \kappa_x & 0 & 0 \\ 0 & \kappa_y & 0 \\ 0 & 0 & \kappa_z \end{pmatrix} = \begin{pmatrix} \dfrac{2\kappa_1\kappa_2}{\kappa_1+\kappa_2} & 0 & 0 \\ 0 & \dfrac{\kappa_1+\kappa_2}{2} & 0 \\ 0 & 0 & \dfrac{\kappa_1+\kappa_2}{2} \end{pmatrix} \quad (3)$$

From (3) above, the longitudinal thermal conductivity ($\kappa_x = \dfrac{2\kappa_1\kappa_2}{\kappa_1+\kappa_2}$) is always less than or equal to the transverse thermal conductivity, ($\kappa_y = \dfrac{\kappa_1+\kappa_2}{2}$), as the harmonic mean is less than or equal to the arithmetic mean. In such a formulation, it was assumed that the layer thickness is sufficiently small, which is equivalent to the tenet that a linear temperature gradient can be defined. Considering a net temperature gradient $\nabla T \left( = \dfrac{T_h - T_c}{l} \right)$ along the total length of the unit comprising layers 1 and 2, the temperature gradient across the individual layers, were derived to be $\nabla T_1 \left( = \dfrac{2\kappa_2 \nabla T}{\kappa_1 + \kappa_2} \right)$ and $\nabla T_2 \left( = \dfrac{2\kappa_1 \nabla T}{\kappa_1 + \kappa_2} \right)$, respectively. We have assumed that the $n$ (an even integer) layers are constituted from a unit cell comprising a layer pair. Consequently, the temperature difference ($T_{dev}$), between that obtained from assuming a linear temperature gradient across the length, (equivalent to defining an effective thermal conductivity for the composite) and that considering temperature variation across the individual layers can be derived to be (also see Supplementary Information, section S2):

$$T_{dev} = \left( \frac{1 - \kappa_2 / \kappa_1}{1 + \kappa_2 / \kappa_1} \right) \frac{l}{2n} \nabla T \qquad (4)$$

It was noted that as $T_{dev}$ decreases as $n$ increases and is proportional to the $\kappa_1 / \kappa_2$ ratio (as indicated in Fig. 2), that approximating the composite medium to possess an equivalent thermal conductivity from (3) is valid for a small layer thickness ($= l/n$) as well as for a higher thermal conductivity contrast, *i.e.,* the approximation is exact for infinitesimally small thickness and when the composite is effectively constituted from a single material, say with $\kappa_2 \rightarrow 0$. Equation (4) above is then a useful relation for understanding the basis for the effective medium approximation.

With the above considerations in mind, consider a horizontal temperature gradient (say, in the *x*-direction) applied to the composite of Fig. 1. When the orientation of the layers in the composite are rotated, in the plane, say, by an angle $\theta$ ($-\pi/2 < \theta < \pi/2$, with $\theta$ considered positive in the counter-clockwise direction – Fig. 3) the layers are now oriented along new axes: *x'* (= *x* *cos θ + y sin θ*) and *y'* (= - *x sin θ + y cos θ*). The originally applied (*horizontal*) temperature gradient now acquires *both* horizontal and vertical components, with respect to the rotated layers. The implication of such a sample re-orientation, for flux manipulation and consequent design of appropriate thermal elements, involves *interpreting* the rotated sample as an *anisotropic* material where the temperature gradient components would be equivalent to the introduction of off-diagonal terms in the thermal conductivity tensor. The key aspect would then be that the degree of anisotropy would change from one composite sample rotation to another and the consequent change may be viewed in terms of *tuning* the $\kappa_{ij}$. Considering that the heat conduction equation, from Equation (1), can be written as $\nabla \cdot (\kappa_{ij} \nabla T_j) = 0$, we can derive that in a changed coordinate system, the *modified* thermal conductivity ($\kappa_{ij}^{\ m}$) would be (as in the Supplementary Information, section S3):

$$\kappa_{ij}^{\ m} = \frac{J \, \kappa_{ij} \, J^T}{\det (J)} \tag{5}$$

where $J$ is the Jacobian matrix of the coordinate transformation between the *new* and the *old* coordinate systems, and for rotation around the *z*-axis: $J = \begin{pmatrix} \cos(\theta) & \sin(\theta) & 0 \\ -\sin(\theta) & \cos(\theta) & 0 \\ 0 & 0 & 1 \end{pmatrix}$, $J^T$ is the transpose of $J$ and det ($J$) denotes the determinant.

Consequently, from the previous inference that such a rotated temperature gradient is equivalent to off-diagonal components in $\kappa_{ij}$, it is implied that the layer rotation has induced anisotropy through the off-diagonal terms in:

$$\kappa_{ij}^{\ m} = \frac{J\kappa_{ij}J^T}{\det(J)} = \begin{pmatrix} \kappa_x \cos^2\theta + \kappa_y \sin^2\theta & (-\kappa_x + \kappa_y)\sin\theta\cos\theta & 0 \\ (-\kappa_x + \kappa_y)\sin\theta\cos\theta & \kappa_y \cos^2\theta + \kappa_x \sin^2\theta & 0 \\ 0 & 0 & \kappa_z \end{pmatrix} \qquad (6)$$

In the transformed coordinates we write the *modified* Fourier law as:

$$q_i^{\ m} = -\kappa_{ij}^{\ m}\nabla T_j^{\ m} \qquad (7)$$

To demonstrate a change in the heat flux density vector, a *unidirectional* temperature gradient (along the line *x'*, – see Fig. 3 for changed axes orientation) is applied to the composite with $\kappa_{ij}^{\ m}$ given by Equation (5). Then,

$$q_i^{\ m} = -[\kappa_x \cos^2\theta + \kappa_y \sin^2\theta]\frac{dT}{dx'}\hat{x}' - [(-\kappa_x + \kappa_y)\sin\theta\cos\theta]\frac{dT}{dy'}\hat{y}' \qquad (8)$$

It is apparent that the heat flux density ($q_i$) has acquired a transverse component, *i.e.,* in the *y'* direction, for $\theta \neq \frac{p\pi}{2}$ where *p* is an integer. The extent of heat flux bending, as inferred through the deviation ($\phi$) of $q_i$ would then be:

$$\phi = Tan^{-1}\left[\frac{(-\kappa_x + \kappa_y)\sin(\theta)\cos(\theta)}{\kappa_x \cos^2(\theta) + \kappa_y \sin^2(\theta)}\right] = Tan^{-1}\left[\frac{(-1+c)\sin(\theta)\cos(\theta)}{\cos^2(\theta) + c\sin^2(\theta)}\right] \qquad (9)$$

In the expression above, on the far right, $c = \kappa_y/\kappa_x = \dfrac{1+\left(\kappa_1\big/\kappa_2\right)^2}{4\cdot\kappa_1\big/\kappa_2}$

We note that $\phi$ (with a period of $\pi$) could be either positive, negative, or zero, which allows us to direct the heat flux density *upwards* or *downwards* through changing/tuning $\theta$ and/or varying $\kappa_1$ and $\kappa_2$. To illustrate the bending of flux lines, we consider the composite in various rotated configurations as indicated in Fig. 3. Here, (a), (b), and (c) represent various sample rotation orientations ($\theta = 0$, $\theta = \pi/4$, and $\theta = -\pi/4$, respectively) while (i), (ii), and (iii) represent the sample geometry, heat flux lines, and temperature variation, respectively. The angle $\theta$ is indicated through the amount of rotation necessary to align the coordinate axes fixed to the sample (*x,y*) *to* the coordinate axes along which the temperature gradient is applied (*x', y'*) and is positive in the counter-clockwise direction. The $\kappa_1$ and $\kappa_2$ of the constituent layers were selected with representative values of 0.1 W/mK and 1 W/mK, respectively, and the surfaces on the right and the left hand side are maintained at 350 K and 300 K, respectively. The end effects and heat loss due to convection and radiation have been ignored. Square geometries (2 cm X 2 cm) were assumed for the simulations (conducted using COMSOL[®] Multiphysics) and *n* was kept constant at 40.

The temperature isotherms on the far right are not orthogonal to the heat flux lines, as represented in the figures in the middle panel. While the bending angle is given through the application of Equation (9), the direction of bending of the flux lines and the corresponding temperature isotherms can also be understood from the viewpoint that while the components of the flux transform covariantly, *i.e.,* in the same way as the basis vectors, the temperature components transform contravariantly. *i.e.,* in the opposite sense to the basis vector change. Additionally, the control of heat flux bending as manifested through $\phi$ was shown to a function of *both* $\theta$ and the $\kappa_1/\kappa_2$ ratio. Fig. 4(a) indicates that $\phi$ initially shows a linear variation with $\kappa_1/\kappa_2$, as a function of $\theta$, and finally saturates at an angle equal to ($90^o$ - $\theta$), as is apparent for larger values of $\theta$ (as also detailed in Supplementary Information, section S4) This is clear from the variation of the last term in Equation (9), which is increasing function of $\kappa_1/\kappa_2$ and approaches a *maximum* value at $\kappa_1/\kappa_2 \rightarrow \infty$, $= Tan^{-1}(Cot(\theta)) = 90 - \theta$. To further understand such a result, we consider for instance, that in the limit $\kappa_1/\kappa_2 \rightarrow \infty$ (see Fig. 1), where with $\kappa_2 \rightarrow 0$, the heat flux would be forced to traverse layer 1, and be bent by an angle of $90^{\circ}$. Alternately, the *minimum value* (with $\kappa_1/\kappa_2 \rightarrow 1$) of the last term in Equation (9) $= Tan^{-1}(0) = 0$. At lower values of $\theta$, the resolution of the horizontal temperature gradient along the rotated layers seems to be less well defined and higher order/cross-coupling effects may be important. The corresponding variation of $\phi$ with $\theta$, at various constant values of $\kappa_1/\kappa_2$ is shown in Fig. 4(b), where the angle of rotation of the composite at which $\phi$ attains maximum ($\theta_\phi^{Max}$) and minimum ($\theta_\phi^{Min}$) was derived to be (as in Supplementary information, section S4):

$$\theta_\phi^{Max} = Tan^{-1}\left[\sqrt{\frac{1}{c}}\right]; \; \theta_\phi^{Min} = -Tan^{-1}\left[\sqrt{\frac{1}{c}}\right] \tag{10}$$

We infer from this figure that a larger contrast in the thermal conductivity between the two layers enables a larger heat flux rotation at a smaller degree of composite rotation, and corresponds to the intuition presented for $\kappa_1/\kappa_2 \rightarrow \infty$.

The implementation of ideas involving anisotropy in appropriate and optimized materials/material configurations, which could then direct the heat propagation *in a different direction* than that in which the thermal gradient is applied is fascinating and could find many applications. We have indicated the underlying issues and guidelines, related to the variation in the heat flux and temperature, and hope that such principles could lead to designer elements for directing thermal energy to useful purpose, *e.g.,* thermal cloaking and concentration.

**Acknowledgments**

The support from the National Science Foundation (Grant ECS 0643761) is acknowledged. Interactions and discussions with R.S. Kapadia are appreciated.

**Figure Captions**

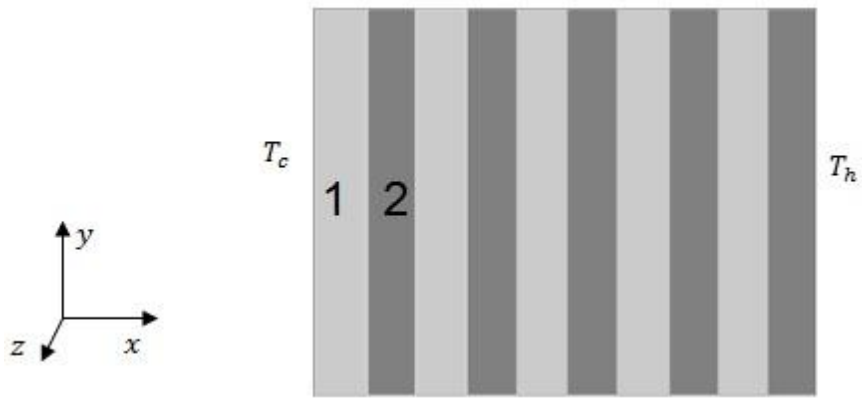

**Figure 1** Illustration of an anisotropic material of total length, *l,* which may be fabricated by alternatively stacking two thin sheets (of individual thickness: $l/n$) – 1 and 2 - of thermal conductivities $\kappa_l$ and $\kappa_2$. The right hand side is maintained at a uniform temperature: $T_h$, while the left hand side is at a lower uniform temperature, $T_c$ ($< T_h$).

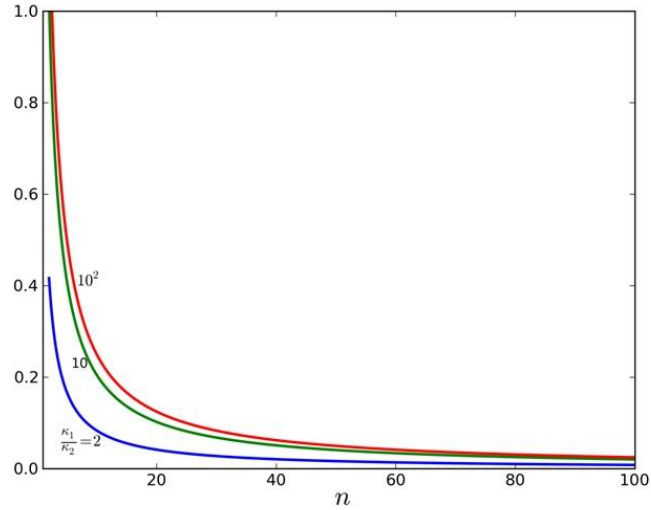

**Figure 2** The average deviation of temperature, $T_{dev}$ from that assumed in an effective medium approximation, plotted as a function of $n$ at various fixed $\kappa_1/\kappa_2$ ratios. The length of the composite $l$ and the temperature gradient were assumed to be 5 cm and 1K/cm, respectively.

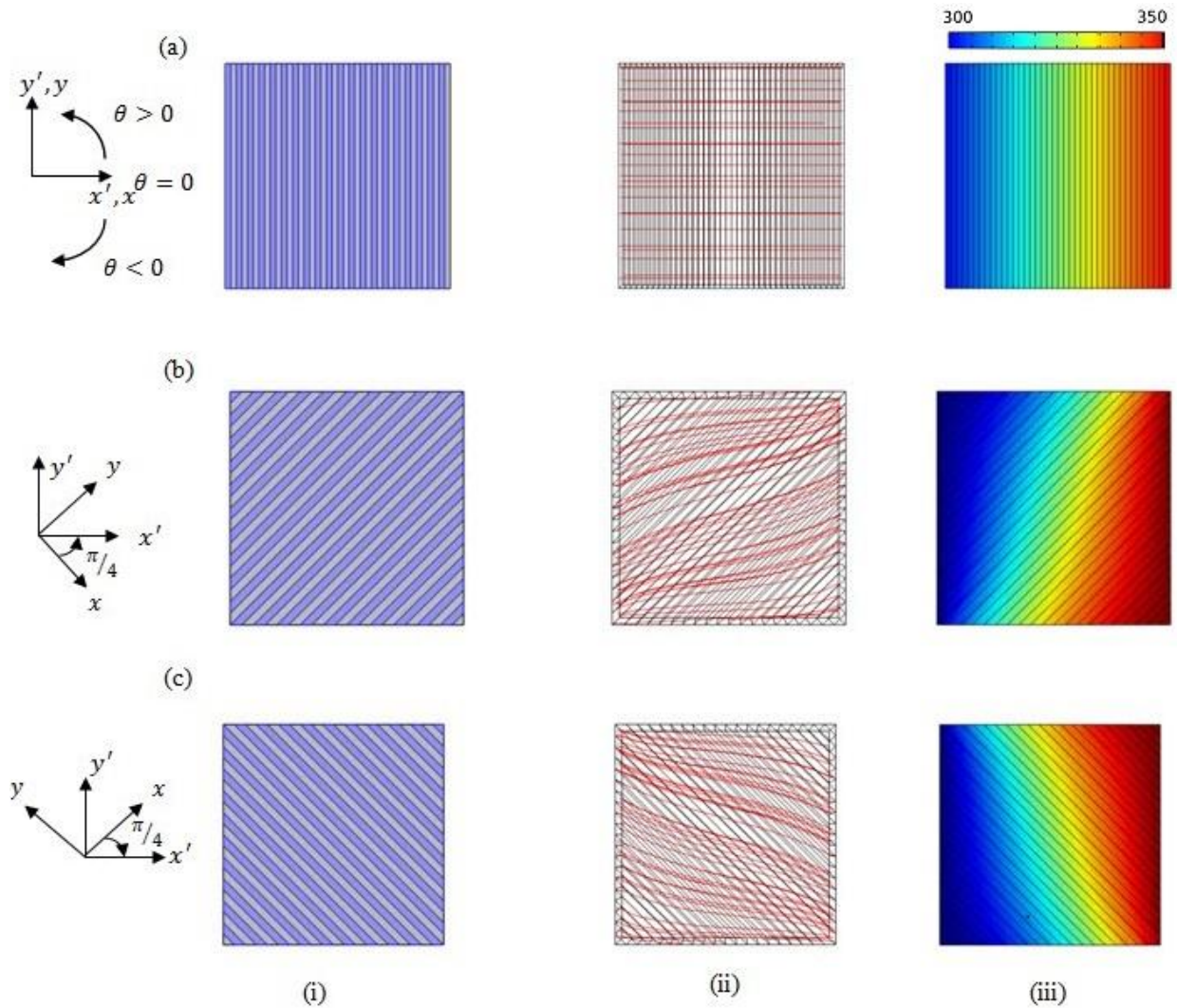

**Figure 3** Heat flux rotation as a function of composite sample orientation. **(a)** the reference layer orientation, with $\theta$ =0, indicates linear propagation of the heat flux and corresponding temperature profile variation, **(b)** when the composite is rotated by $\theta = \pi/4$, a *downwards* bending of the heat flux is indicated, while when the composite is rotated by **(c)** $\theta = -\pi/4$, an *upwards* bending of the heat flux is shown. The bending angle is given through Equation (9). $x, y$ and $x', y'$ correspond to original and rotated coordinate systems respectively.

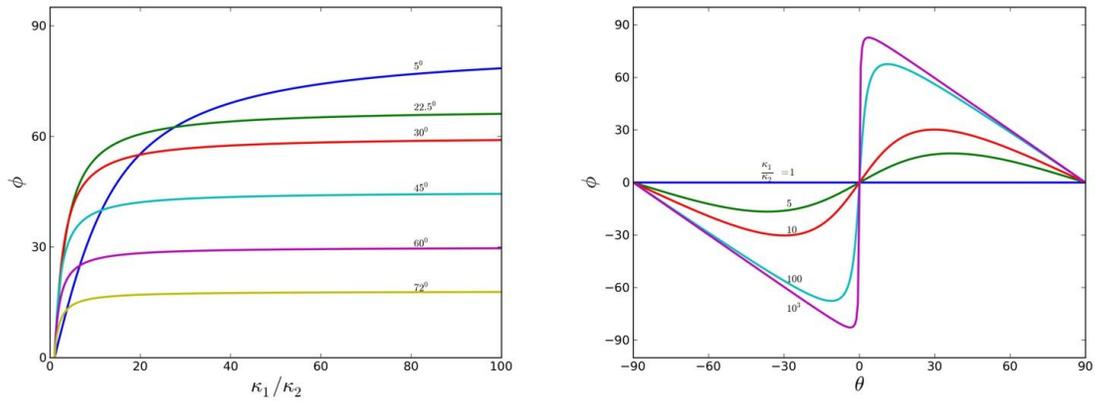

**Figure 4 (a)** The variation of the flux bending angle $\phi$ *vs.* **(a)** $\kappa_1/\kappa_2$ at fixed values of the composite rotation angle, $\theta$, **(b)** $\theta$ at fixed values of $\kappa_1/\kappa_2$.

# <u>Supplementary Information</u>

## S1. Derivation of (3) in the paper

(a) <u>Series thermal conductivity</u>

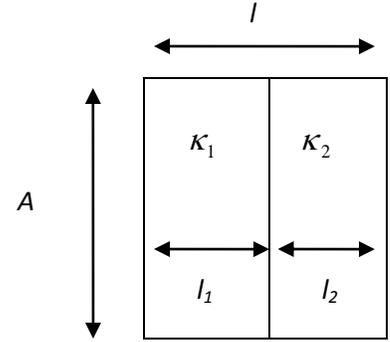

**Figure S1 (a)**

$$q_1 = -\kappa_1 \frac{\Delta T_1}{l_1} A \qquad \text{(Heat flux density in layer1)}$$

$$q_2 = -\kappa_2 \frac{\Delta T_2}{l_2} A \qquad \text{(Heat flux density in layer2)}$$

$$q = -\kappa_s \frac{\Delta T_1 + \Delta T_2}{l_1 + l_2} A \qquad \text{(Total heat flux density in the composite)}$$

($\kappa_s$ is the effective thermal conductivity for the configuration , $\Delta T_1$ & $\Delta T_2$ are temperature gradients across layer (1) and layer (2) respectively in Figure S1 (a))

As the heat flux is continuous across the layers in series, $q = q_1 = q_2$

Solving for $\kappa_s$, we get $\kappa_s = \frac{(l_1 + l_2)\kappa_1\kappa_2}{l_1\kappa_2 + l_2\kappa_1}$. In **(3)**, it is assumed that $l_1 = l_2$

(b) <u>Parallel thermal conductivity</u>

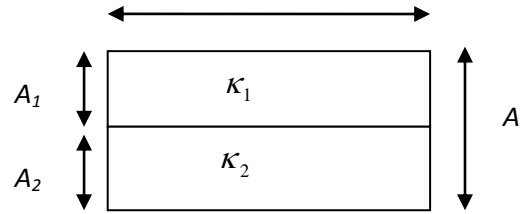

**Figure S1 (b)**

$$q_1 = -\kappa_1 \frac{\Delta T}{l} A_1 \quad \text{(Heat flux density in layer1)}$$

$$q_2 = -\kappa_2 \frac{\Delta T}{l} A_2 \quad \text{(Heat flux density in layer2)}$$

Total heat flux $q = -\kappa_p \frac{\Delta T}{l}(A_1 + A_2)$ (Total heat flux density in the composite)

($\kappa_p$ is the effective thermal conductivity for the configuration, $\Delta T$ is the temperature gradient across composite in Figure S1 (b))

As the layers are in parallel configuration, $q = q_1 + q_2$

Solving for $\kappa_p$, we get: $\kappa_p = \frac{\kappa_1 A_1 + \kappa_2 A_2}{A_1 + A_2}$. In **(3)**, it is assumed that $A_1 = A_2 = A/2$

**S2. Temperature deviation from the effective medium approximation, related to Figure 2**

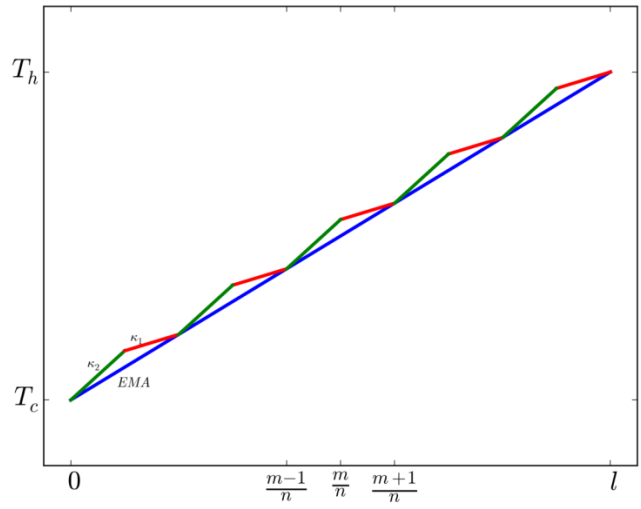

**Figure S2** Comparison of the simulated temperature profiles in $\kappa_2$ (green) and $\kappa_l$ (red) to the temperature profile assuming EMA (Blue line) for the composite in Figure 2.

$$T_{dev} = \frac{\text{Total area enclosed by the triangles}}{\text{Length of the composite}(l)}$$

$$T_{dev}\left(\frac{\kappa_1}{\kappa_2},\right) = \left(\frac{1 - \frac{\kappa_2}{\kappa_1}}{1 + \frac{\kappa_2}{\kappa_1}}\right)\frac{l}{2n}\nabla T$$

This is **equation (4)** in the paper

## S3. The modification of thermal conductivity on coordinate transformation

*From the heat conduction equation:* $\rho \, C \dfrac{\partial T}{\partial t} + \nabla.(-\kappa \nabla T) = Q$

where, $\rho$ is the density, $C$ is the heat capacity, $T$ is the temperature, $\kappa$ is the thermal conductivity tensor and $Q$ is the heat source.

At steady state, $\nabla.(-\kappa \nabla T) = \dfrac{-\partial}{\partial x_i} \kappa_{ij} \dfrac{\partial T}{\partial x_j} = Q$

Multiplying on the left by $\dfrac{\partial x'_a}{\partial x'_a}$ and $\dfrac{\partial x'_b}{\partial x'_b}$ and on the right, and simplifying, we get:

$$\dfrac{-\partial}{\partial x_i} \dfrac{\partial x'_a}{\partial x'_a} \kappa_{ij} \dfrac{\partial T}{\partial x_j} \dfrac{\partial x'_b}{\partial x'_b} = Q$$

Rearranging,

$$\dfrac{-\partial}{\partial x'_a} \dfrac{\partial x'_a}{\partial x_i} \kappa_{ij} \dfrac{\partial x'_j{}^{T}}{\partial x_b} \dfrac{\partial T}{\partial x'_b} = Q \qquad (\dfrac{\partial x'_j{}^{T}}{\partial x_b} = \dfrac{\partial x'_b}{\partial x_j})$$

yielding, $\dfrac{-\partial}{\partial x'_a} A_{aj} \kappa_{ij} A_{jb}{}^{T} \dfrac{\partial T}{\partial x'_b} = Q$

(where, $A_{aj} = \dfrac{\partial x'_a}{\partial x_i}$ and $A_{jb}{}^{T} = \dfrac{\partial x'_j{}^{T}}{\partial x_b}$ , and $A$ is the Jacobian matrix of the coordinate

transformation from $[x_1, x_2, x_3]$ to $[x'_1(x_1, x_2, x_3), x'_2(x_1, x_2, x_3), x'_3(x_1, x_2, x_3)]$

with $A_{ij} = \dfrac{\partial x'_i}{\partial x_j}$; $\nabla = \dfrac{\partial}{\partial x_j} e_j$; $\nabla' = \dfrac{\partial}{\partial x'_j} e_j$

Then, $\dfrac{1}{\det A} \dfrac{-\partial}{\partial x'_b} A_{aj} \kappa_{ij} A_{jb}{}^{T} \dfrac{\partial T}{\partial x'_b} = \dfrac{Q}{\det A}$

$\nabla. \left[ \dfrac{-A \kappa A^{T}}{\det A} \nabla' T \right] = Q'$ where, $Q' = \dfrac{Q}{\det A}$ preserves the form of the heat conduction equation

Finally, we get **equation (5)** of the paper

$\nabla'.[-\kappa' \nabla' T] = Q'$ ($\kappa = \dfrac{A \kappa A^{T}{}'}{\det A}$ ; the thermal conductivity tensor in transformed coordinates)

**S4. <u>Heat flux bending angle, $\phi$ at a fixed $\theta$</u>**

From **equation (9)** in the paper:

$$\phi = Tan^{-1}\left[\frac{(-\kappa_x + \kappa_y)\sin(\theta)\cos(\theta)}{\kappa_x\cos^2(\theta) + \kappa_y\sin^2(\theta)}\right]$$

$$\kappa_x = \frac{2\kappa_1\kappa_2}{\kappa_1 + \kappa_2} \quad, \qquad \kappa_y = \frac{\kappa_1 + \kappa_2}{2}$$

Let $a = \kappa_1/\kappa_2$ and $c(a) = \kappa_y/\kappa_x = \dfrac{(1+a)^2}{4a}$

As $c(a)$ is symmetric with respect to $a$ and $1/a$, we consider $a > 1$. Then,

$$\phi = Tan^{-1}\left[\frac{(-1+c)\sin(\theta)\cos(\theta)}{\cos^2(\theta) + c\sin^2(\theta)}\right]$$

Assuming that $\theta$ is a constant, $\phi = f(c)$

$$f(c) = Tan^{-1}\left[\frac{(-1+c)\sin(\theta)\cos(\theta)}{\cos^2(\theta) + c\sin^2(\theta)}\right] \qquad \text{with} \qquad g(c) = \frac{(-1+c)\sin(\theta)\cos(\theta)}{\cos^2(\theta) + c\sin^2(\theta)}$$

$$f(c) = Tan^{-1}\big[g(c)\big]$$

<u>Characteristics</u>

1) $f(c)$ is an increasing function

2) $g(c)$, $c(a)$ are increasing functions. Thus $g(a)$ is an increasing function.

3) Max value of $a$ can be infinity, min value can be 1. (As $c$ is symmetric to $a$ and $1/a$)

4) Thus, $f(c) = Tan^{-1}(g(a))$ is an increasing function of $a$

5) We are interested in finding:

    a. *maximum* of the function (As $a \rightarrow \infty$, $c \rightarrow$ zero $\Rightarrow g(c)$ tends to $cot\ \theta$

        $\Rightarrow$ maximum value of $f(c) = Tan^{-1}(Cot(\theta)) = 90 - \theta$

    b. *minimum* of the function (As $a \rightarrow 1$, $c \rightarrow \infty$ $\Rightarrow g(c)$ tends to 0

        $\Rightarrow$ minimum of $f(c) = Tan^{-1}(0) = 0$

($\theta \neq n\pi$, and $c \neq 1$ as this would correspond to isotropic material and $\phi$ cannot be zero).

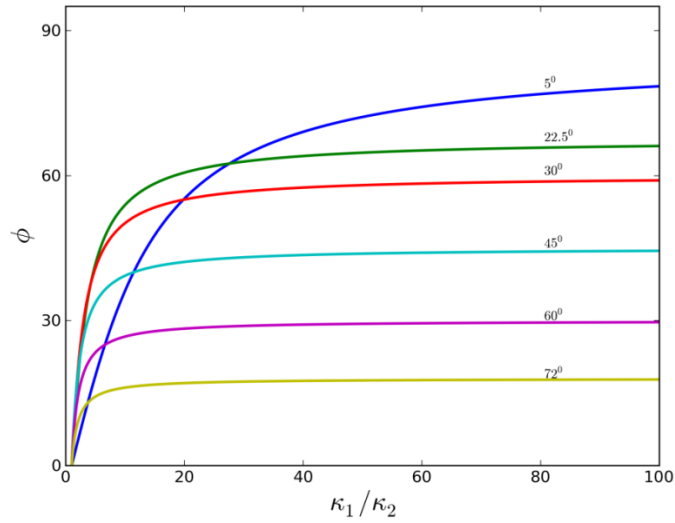

**Figure S4 (a)** Variation of $\phi$ *vs* $\kappa_1 / \kappa_2$, for different values of $\theta$

## Heat flux bending angle, $\phi$ for a fixed $\kappa_1 / \kappa_2$

Assuming that $a = \kappa_1 / \kappa_2$ is fixed and $\theta$ is varying,

$$h(\theta) = Tan^{-1}\left[\frac{(-1+c)\sin(\theta)\cos(\theta)}{\cos^2(\theta) + c\sin^2(\theta)}\right]$$

$$\frac{\partial h}{\partial \theta} = \frac{(-1+c)(\cos^2(\theta) - c\sin^2(\theta))}{\cos^2(\theta) + c^2\sin^2(\theta)}$$

The value of $\theta$ for *maximum* and *minimum* $\phi$ : $\theta = \pm Tan^{-1}\left[\sqrt{\dfrac{1}{c}}\right]$  -  **equation (10)** in the paper

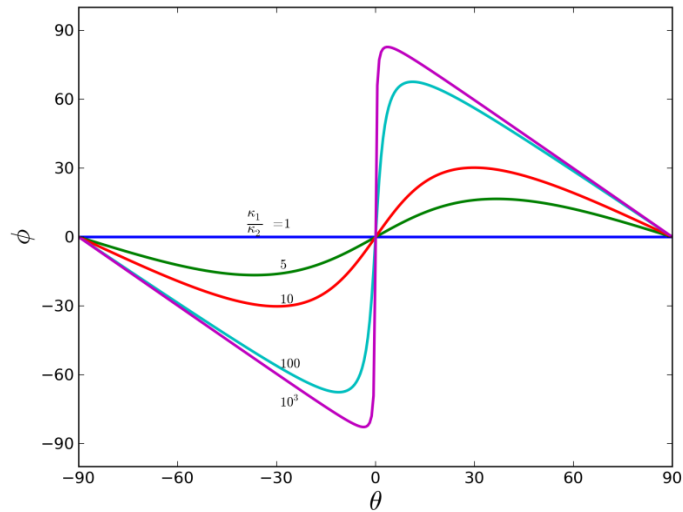

**Figure S4 (b)** Variation of $\phi$ *vs* $\theta$, for different $a = \kappa_1 / \kappa_2$